\pdfoutput=1

\documentclass[twocolumn,english,pra,floatfix,showpacs,showkeys]{revtex4-1}
\usepackage[T1]{fontenc}
\usepackage{amsmath}
\usepackage{babel}
\usepackage{graphicx}
\usepackage{subfig}

\begin{document}

\title{Correlation function and mutual information}

\author{R.X. Dong}
\affiliation{Institute of Physics, Chinese Academy of Sciences, Beijing 100190, China}

\author{D.L. Zhou}
\affiliation{Institute of Physics, Chinese Academy of Sciences, Beijing 100190, China}

\begin{abstract}
Correlation function and mutual information are two powerful tools
to characterize the correlations in a quantum state of a composite
system, widely used in many-body physics and in quantum information
science, respectively. We find that these two tools may give different conclusions about the order of the degrees of correlation
in two specific two-qubit states. This result implies that the orderings of bipartite quantum states according to the degrees of correlation  depend on which correlation measure we adopt.
\end{abstract}



\maketitle

\section{Introduction}

Entanglement is a kind of correlation in a quantum state of a composite
system, which can not be simulated only with classical resources \cite{Wer89}.
The research on entanglement has become a rich branch in the research
of quantum information science \cite{HHHH09}. Comparing with entanglement,
correlation functions have been proved to be useful concepts in describing
the correlation effects in many-body physics. Since the correlation
functions \cite{Ma85} are directly related with observables, they
find many successful applications in many-body physics, e.g. correlation
is a crucial element for understanding critical phenomena \cite{Wil75}. 

A basic idea is to distinguish different types of correlations in
a multipartite quantum state. For example, correlations in a quantum
state can be classified into classical correlation and quantum correlation,
and entanglement is contained in quantum correlation \cite{HV02,OZ02,GPW05,Luo08,Wal09,MPSVW10}.
In quantum information science, the degree of (total) correlation
in a composite quantum state is usually described mathematically by
mutual information. Notice that correlation functions do not distinguish
classical or quantum correlations either, correlation functions and
mutual entropy describe the same (or similar) properties of the quantum
state. Intuitively speaking, the larger the magnitude of the correlation function or the mutual entropy for a quantum state is, the more correlated the state is. Then the ordering of quantum states on the degrees of correlation can be built by comparing the magnitude of  the correlation functions or the mutual entropy. A natural question is whether they give the same ordering of
quantum states on the degrees of correlation.

This question is reminiscent of a similar situation when dealing with
entanglement. The general concept of entanglement monotone is introduced
to characterize the order on the degree of entanglement \cite{Vid00}.
Different entanglement monotones may give different ordering of quantum
states on the degrees of entanglement \cite{EP99,MG04}. This consideration
motivates us to take a similar way to solve the present problem on
correlation instead of entanglement.

In this article, we will perform a comparative study on the correlation
functions and mutual information in characterizing the degree of correlation.
This article is organized as follows. In Sec. II, we will briefly
review two correlation measures, where the first one is directly related
with the correlation functions. In Sec. III, we will compare the above
two correlation measures on the classically correlated two-qubit states
. In Sec. IV, we will give three relative discussions on the problem.
Finally, we will give a brief summary.

\section{Two correlation measures}

We consider a quantum state $\rho^{(12)}$ for a system composed by two subsystems $1$ and $2$. A correlation measure for the bipartite quantum state $\rho^{(12)}$,
denoted as $C(\rho^{(12)})$, is a function that should satisfy four
basic requirements \cite{HV02,ZZXY06}: 
\begin{enumerate}
\item It is semi-positive, i.e. $C(\rho^{(12)})\ge0$. 
\item It is zero if and only if the two partite state $\rho^{(12)}$ is
a product state, i.e. $C(\rho^{(12)})=0$ $\Leftrightarrow$ $\rho^{(12)}=\rho^{(1)}\otimes\rho^{(2)}$. 
\item It is invariant under local unitary transformations, i.e. $C(U^{(1)}U^{(2)}\rho^{(12)}U^{(2)\dagger}U^{(1)\dagger})=C(\rho^{(12)})$. 
\item It does not increase under local operations, i.e. $C(\mathcal{E}^{(1)}\mathcal{E}^{(2)}(\rho^{(12)}))\le C(\rho^{(12)})$. 
\end{enumerate}
These four requirements, however, does not imply that there exists
a unique correlation measure.

A possible way to construct a correlation measure is given as follows.
For any given bipartite quantum state $\rho^{(12)}$, we can introduce
a product state $\tilde{\rho}^{(12)}=\rho^{(1)}\otimes\rho^{(2)}$,
where $\rho^{(1)}$ and $\rho^{(2)}$ are reduced density matrices
of $\rho^{(12)}$ . Then we use different distance measures to characterize
the difference between $\rho^{(12)}$ and $\tilde{\rho}^{(12)}$.
By this way, we construct two correlation measures for a bipartite
quantum state $\rho^{(12)}$. The first one \cite{ZZXY06} is \begin{equation}
C_{I}(\rho^{(12)})=D(\rho^{(12)},\tilde{\rho}^{(12)}),\end{equation}
 where the trace distance $D(\sigma,\tau)\equiv\frac{1}{2}\mathrm{\mathrm{Tr}}\vert\sigma-\tau\vert$
\cite{NC00}.

The second one is \begin{equation}
C_{II}(\rho^{(12)})=S(\rho^{(12)}\vert\vert\tilde{\rho}^{(12)}),\end{equation}
 where the quantum relative entropy $S(\sigma\vert\vert\tau)\equiv\mathrm{Tr}(\sigma(\ln\sigma-\ln\tau))$
is a distance-like function \cite{NC00}.

It is easy to prove that the two functions $C_{I}(\rho^{(12)})$ and
$C_{II}(\rho^{(12)})$ satisfy the four basic requirements of a correlation
measure, and thus both of them are legitimate correlation measures.
A natural question arises: do these correlation measures give the
same ordering on the degrees of correlations for all bipartite quantum
states? Here the same order means the following equivalence relation:
for any bipartite quantum states $\rho^{(12)}$ and $\sigma^{(12)}$,
\begin{equation}
C_{I}(\rho^{(12)})>C_{I}(\sigma^{(12)})\Longleftrightarrow C_{II}(\rho^{(12)})>C_{II}(\sigma^{(12)}).\label{c1c2so}\end{equation}
 The basic requirements of a correlation measure ensure that $C_{I}(\rho^{(12)})$
and $C_{II}(\rho^{(12)})$ will give the same order of correlation
for some classes of bipartite quantum states. For example, if a
bipartite state can be obtained by local operations on the other
state, then all the measures of correlation will give the same order
for these two states according to the fourth requirement.

Before investigating the above problem in detail, we first build a
direct relation between correlation functions and the correlation
measure $C_{I}$.

\subsection*{Correlation functions and $C_{I}$}

To answer the question raised above, we find that it is sufficient to restrict our discussions to the two-qubit
case. For a two-qubit system, the correlation function is defined
by \begin{equation}
C_{F}(\sigma_{i}^{(1)},\sigma_{j}^{(2)})=\mathrm{Tr}(\rho^{(12)}\sigma_{i}^{(1)}\sigma_{j}^{(2)})-\mathrm{Tr}(\rho^{(1)}\sigma_{i}^{(1)})\mathrm{\mathrm{Tr}(\rho^{(2)}\sigma_{j}^{(2)}),}\end{equation}
 where $i,j\in\{x,y,z\}$. We observe that, on one hand, a correlation
function $C_{F}(\sigma_{i}^{(1)},\sigma_{j}^{(2)})$ does not satisfy
any basic requirement of a correlation measure, and it is not a legitimate
correlation measure. On the other hand, if a two partite state is
a product state, then the correlation function is zero. In this sense,
a correlation function is the witness of correlation. Notice that
\begin{equation}
\rho^{(12)}-\rho^{(1)}\rho^{(2)}=\frac{1}{4}\sum_{i,j}C_{F}(\sigma_{i}^{(1)},\sigma_{j}^{(2)})\sigma_{i}^{(1)}\sigma_{j}^{(2)}.\label{corop}\end{equation}
 Eq.(\ref{corop}) implies that the information of all the independent
correlation functions are contained in the operator $\rho^{(12)}-\rho^{(1)}\rho^{(2)}$,
which completely determines the correlation measure $C_{I}(\rho^{(12)})$.
Therefore, the correlation measure $C_{I}(\rho^{(12)})$ is directly
related with all the independent correlation functions.

\section{Comparisons of correlation measures}

To compare the correlation measures $C_{I}(\rho^{(12)})$ and $C_{II}(\rho^{(12)})$,
we consider a classically correlated two-qubit state \cite{Luo08,MPSVW10}
\begin{equation}
\rho^{(12)}=p_{00}\vert00\rangle\langle00\vert+p_{01}\vert01\rangle\langle01\vert+p_{10}\vert10\rangle\langle10\vert+p_{11}\vert11\rangle\langle11\vert,\label{rhop}\end{equation}
 where $0\le p_{00},p_{01},p_{10},p_{11}\le1$ and $p_{00}+p_{01}+p_{10}+p_{11}=1$.

A direct calculation gives that \begin{equation}
C_{I}(\rho^{(12)})=2\vert p_{00}p_{11}-p_{01}p_{10}\vert,\label{c1p}\end{equation}
 and \begin{eqnarray}
C_{II}(\rho^{(12)}) & = & -(p_{00}+p_{01})\ln(p_{00}+p_{01})-(p_{10}+p_{11})\nonumber \\
 &  & \times\ln(p_{10}+p_{11})-(p_{00}+p_{10})\ln(p_{00}+p_{10})\nonumber \\
 &  & -(p_{01}+p_{11})\ln(p_{01}+p_{11})+p_{00}\ln p_{00}\nonumber \\
 &  & +p_{01}\ln p_{01}+p_{10}\ln p_{10}+p_{11}\ln p_{11}.\label{c2p}\end{eqnarray}

For the two-qubit state (\ref{rhop}), the unique nonzero correlation
function is $C_{F}(\sigma_{z}^{(1)},\sigma_{z}^{(2)})$, and we have
\begin{equation}
C_{I}(\rho^{(12)})=\frac{1}{2}\vert C_{F}(\sigma_{z}^{(1)},\sigma_{z}^{(2)})\vert.\end{equation}

Eq. (\ref{c1p}) and (\ref{c2p}) show that the mathematical expressions
for the two correlation measures are very different, and the relation
between them is not simple. What are the similarities and differences
between the two correlation measures? We will compare these two measures
by numerical and analytical methods in the following subsections.

\subsection{Numerical results}

Since the two correlation measures $C_{I}$ and $C_{II}$ are the
functions of three independent probabilities, we had better scan one
probability to make they can be demonstrated in a $2$-dimensional
figure. First, we scan the probability $p_{10}$, and show the results
in Fig. 1.

\begin{figure}[ht]
 \centering 
 \captionsetup[subfloat]{nearskip=-3pt} \subfloat[$C_{I}$
when $p_{10}=0.1$.]{ \includegraphics[width=0.24\textwidth]{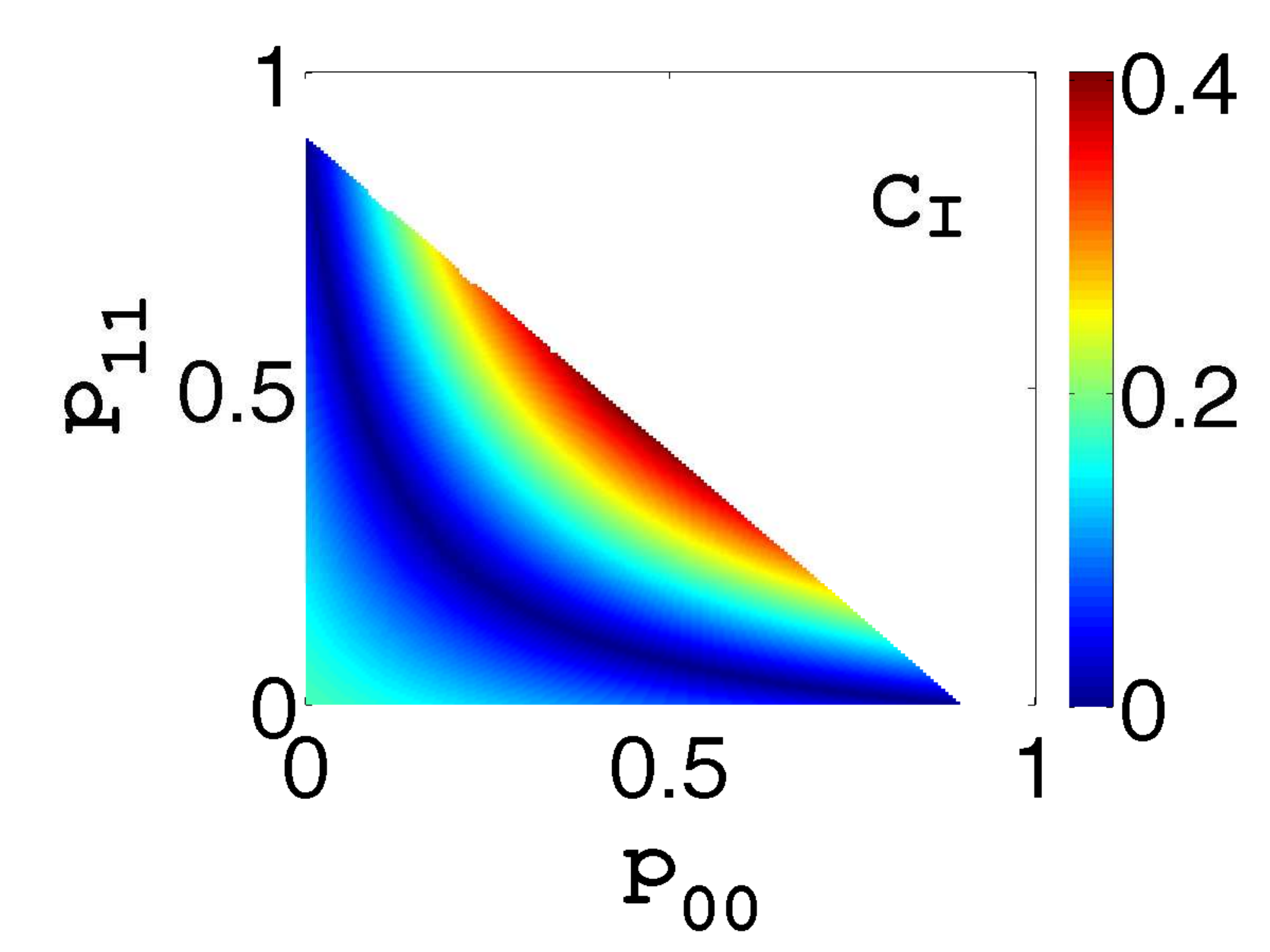}
\label{fig1:1:1} } \subfloat[$C_{II}$ when $p_{10}=0.1$.]{
\includegraphics[width=0.24\textwidth]{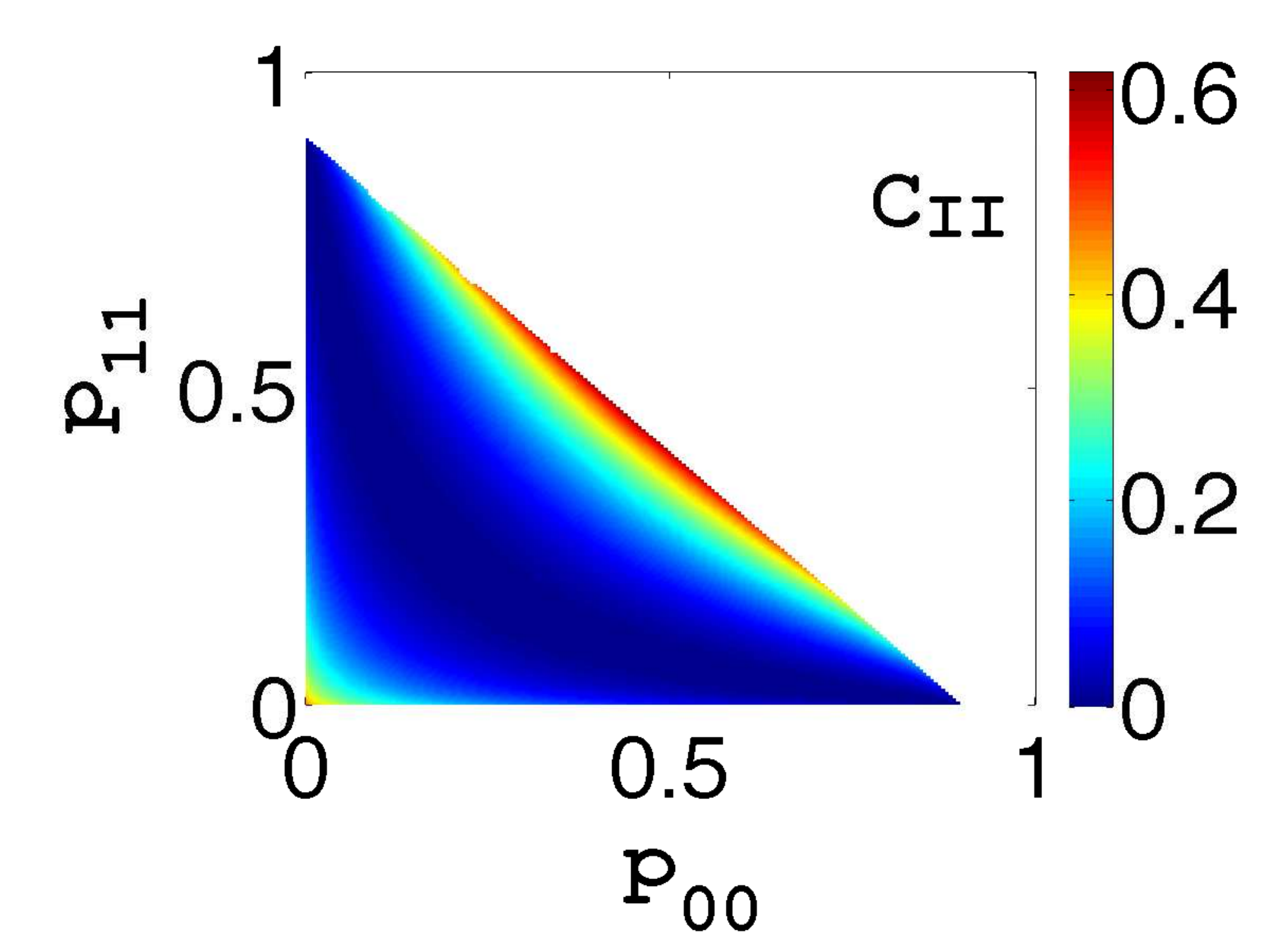}
\label{fig1:1:2} } \\
 \subfloat[$C_{I}$ when $p_{10}=0.4$.]{ \includegraphics[width=0.24\textwidth]{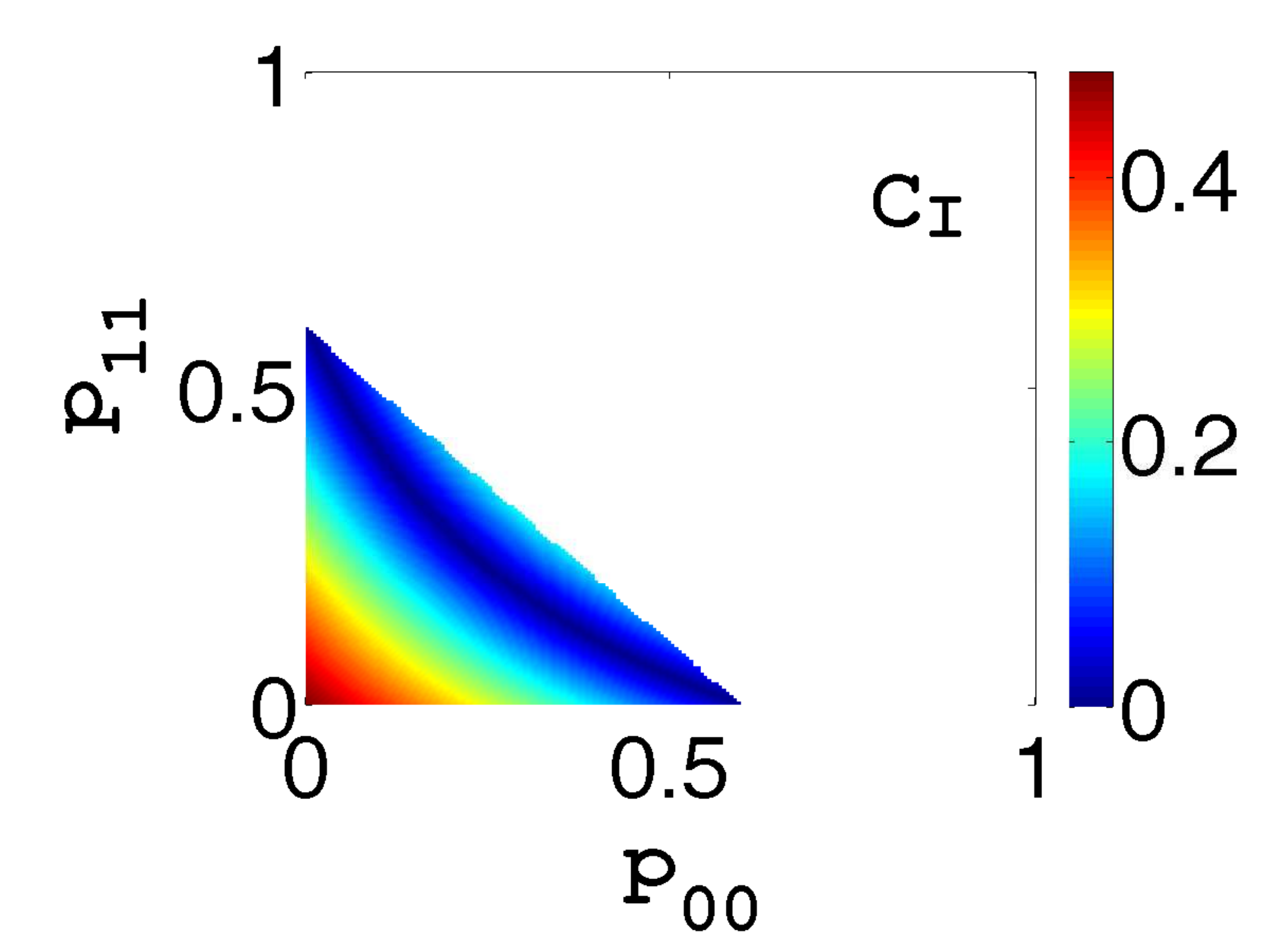}
\label{fig1:2:1} } \subfloat[$C_{II}$ when $p_{10}=0.4$.]{
\includegraphics[width=0.24\textwidth]{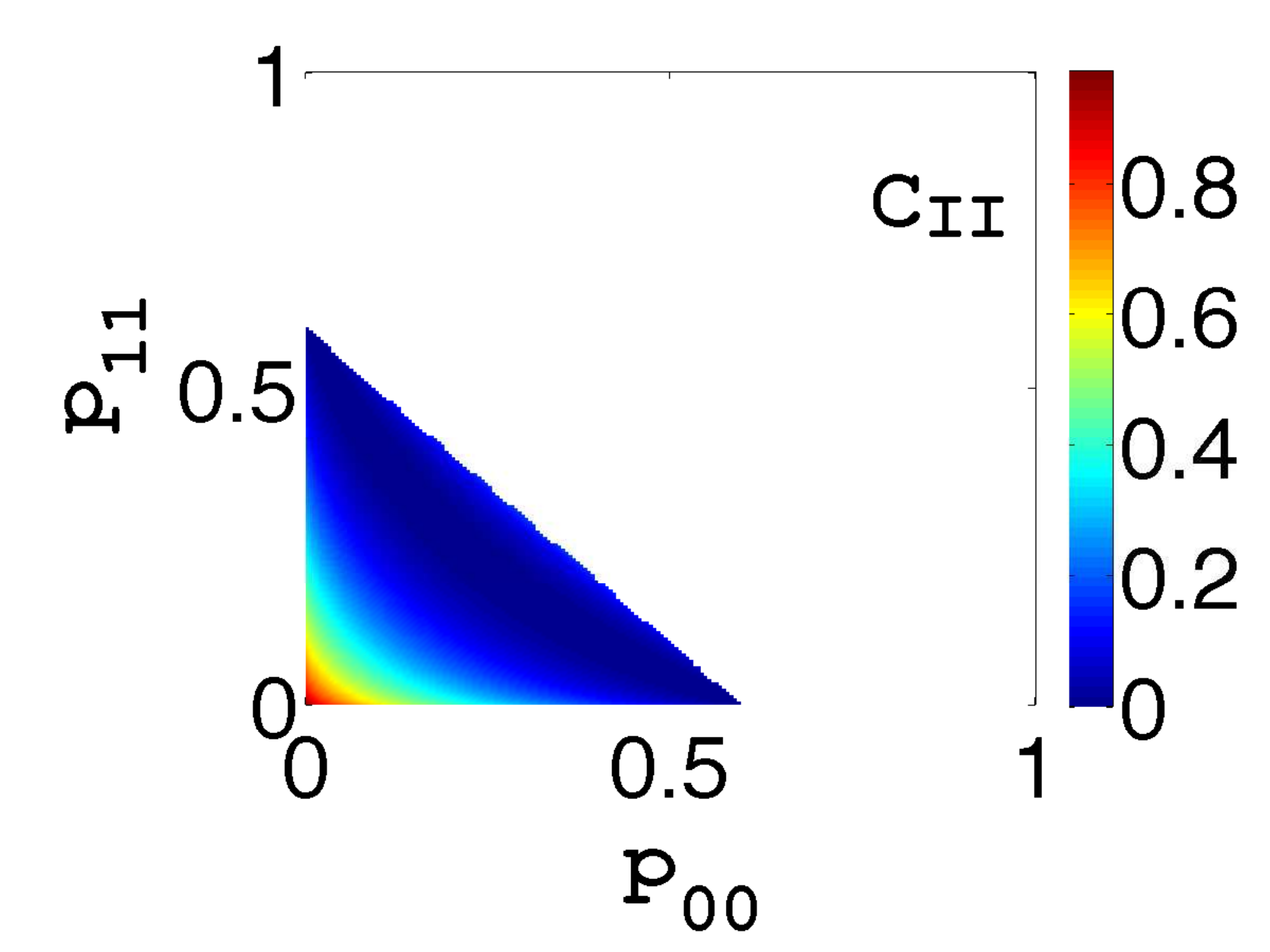}
\label{fig1:2:2} } \\
 \subfloat[$C_{I}$ when $p_{10}=0.7$.]{ \includegraphics[width=0.24\textwidth]{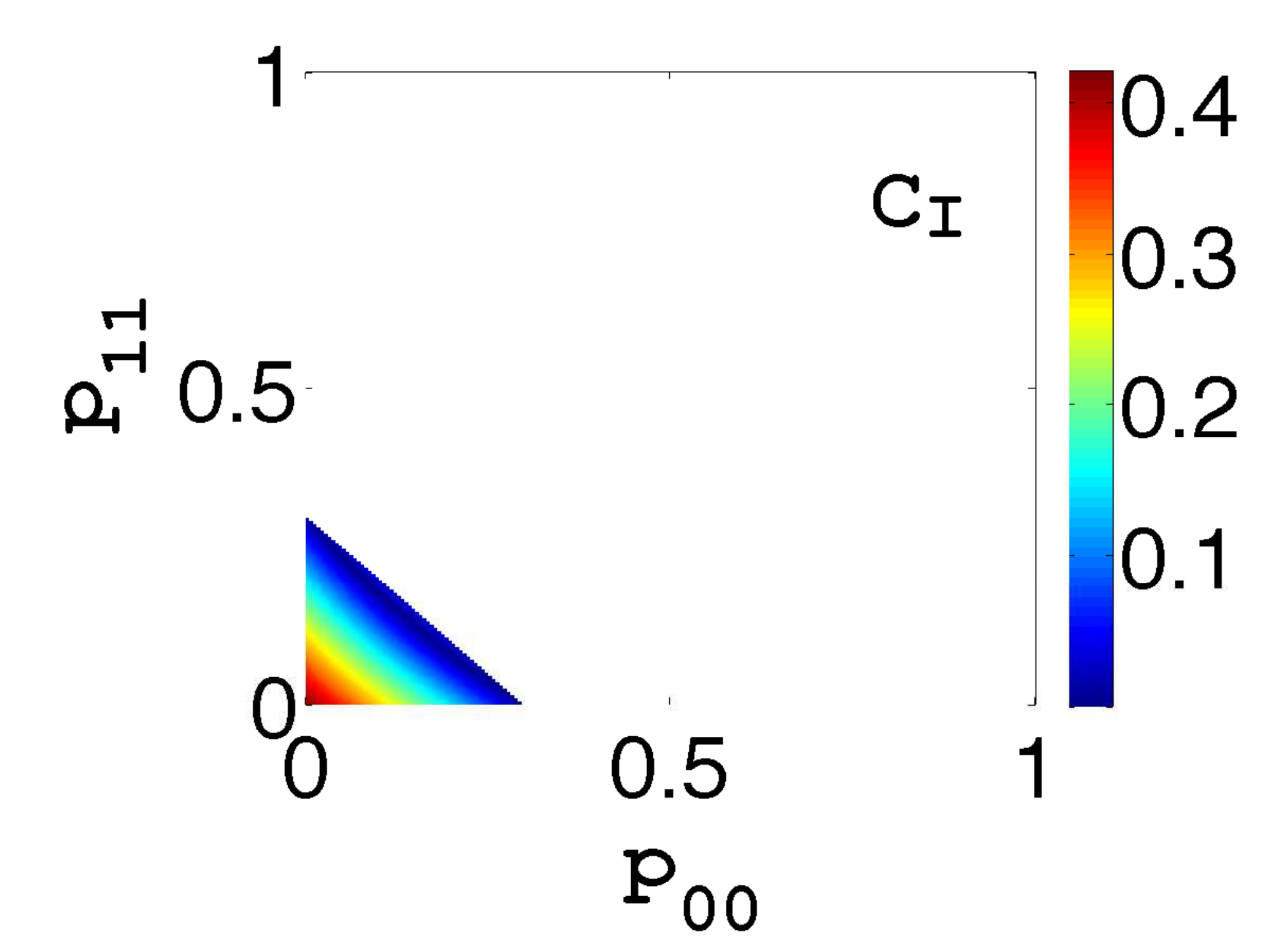}
\label{fig1:3:1} } \subfloat[$C_{II}$ when $p_{10}=0.7$.]{
\includegraphics[width=0.24\textwidth]{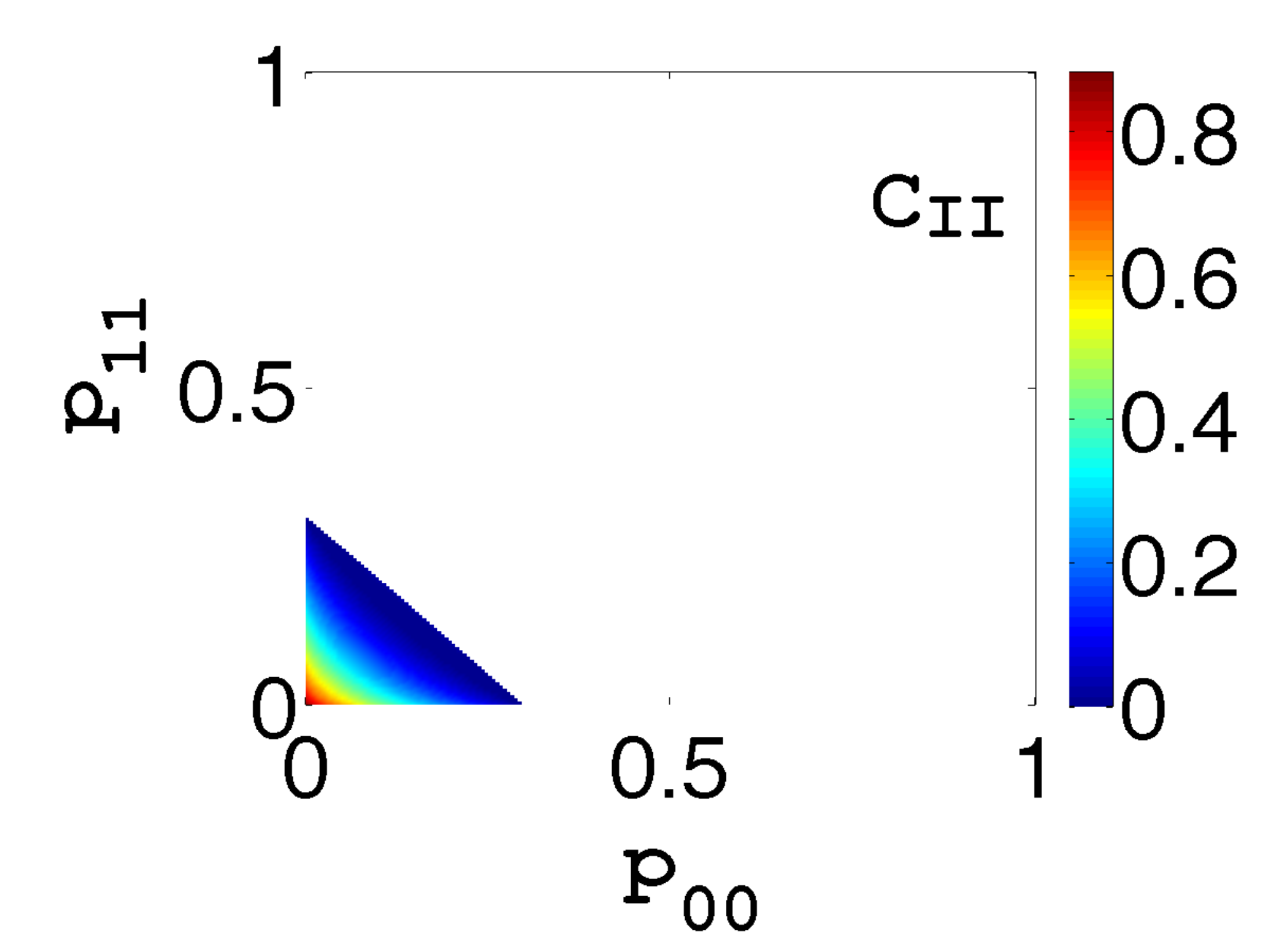}
\label{fig1:3:2} } \caption{Numerical comparison of $C_{I}$ and $C_{II}$ with varying $p_{10}$.}

\label{fig1} 
\end{figure}

The numerical results in Fig. 1 show that these two measures give
very similar behaviors. More precisely, $C_{I}$ and $C_{II}$ show
the same increasing or decreasing behavior with varying the parameters.

To demonstrate the universality of the above observation, we further
scan the probability $p_{11}$ to obtain the two correlation measures
in Fig. 2. We find a similar behavior of the two correlation measures
as those appeared in Fig. 1.

\begin{figure}[ht]
 \centering \captionsetup[subfloat]{nearskip=-3pt} \subfloat[$C_{I}$
when $p_{11}=0.1$.]{ \includegraphics[width=0.24\textwidth]{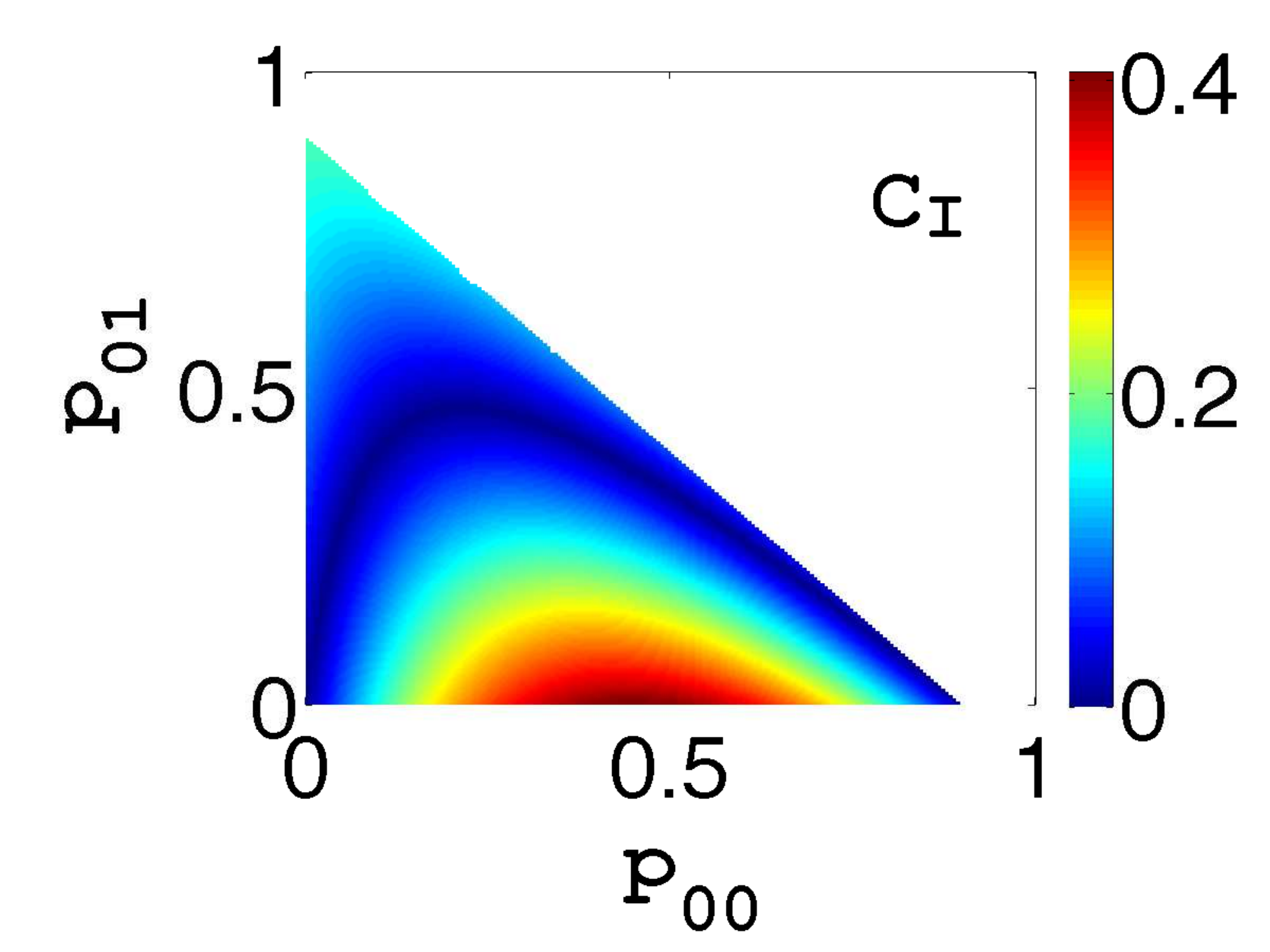}
\label{fig2:1:1} } \subfloat[$C_{II}$ when $p_{11}=0.1$.]{
\includegraphics[width=0.24\textwidth]{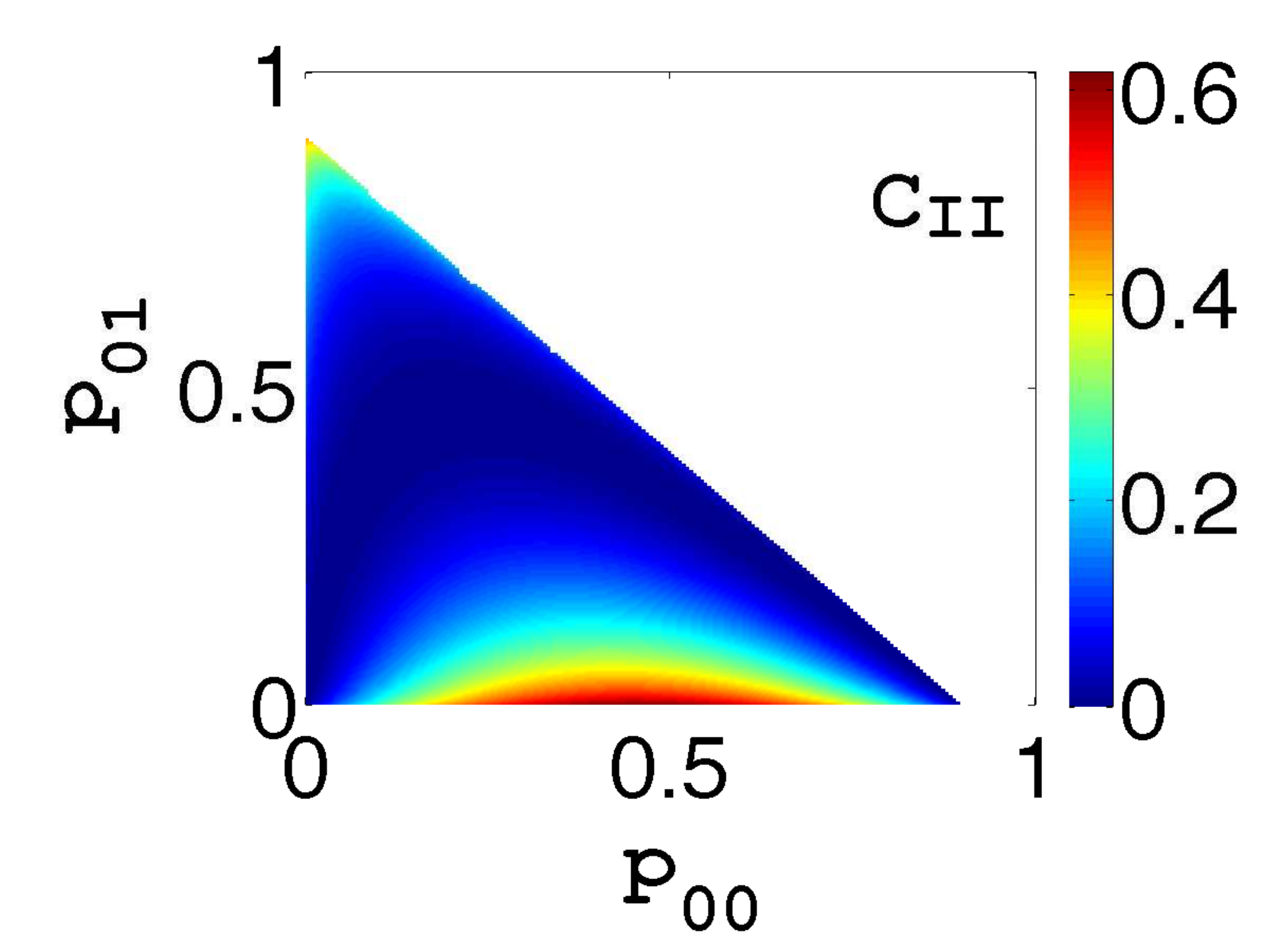}
\label{fig2:1:2} } \\
 \subfloat[$C_{I}$ when $p_{11}=0.4$.]{ \includegraphics[width=0.24\textwidth]{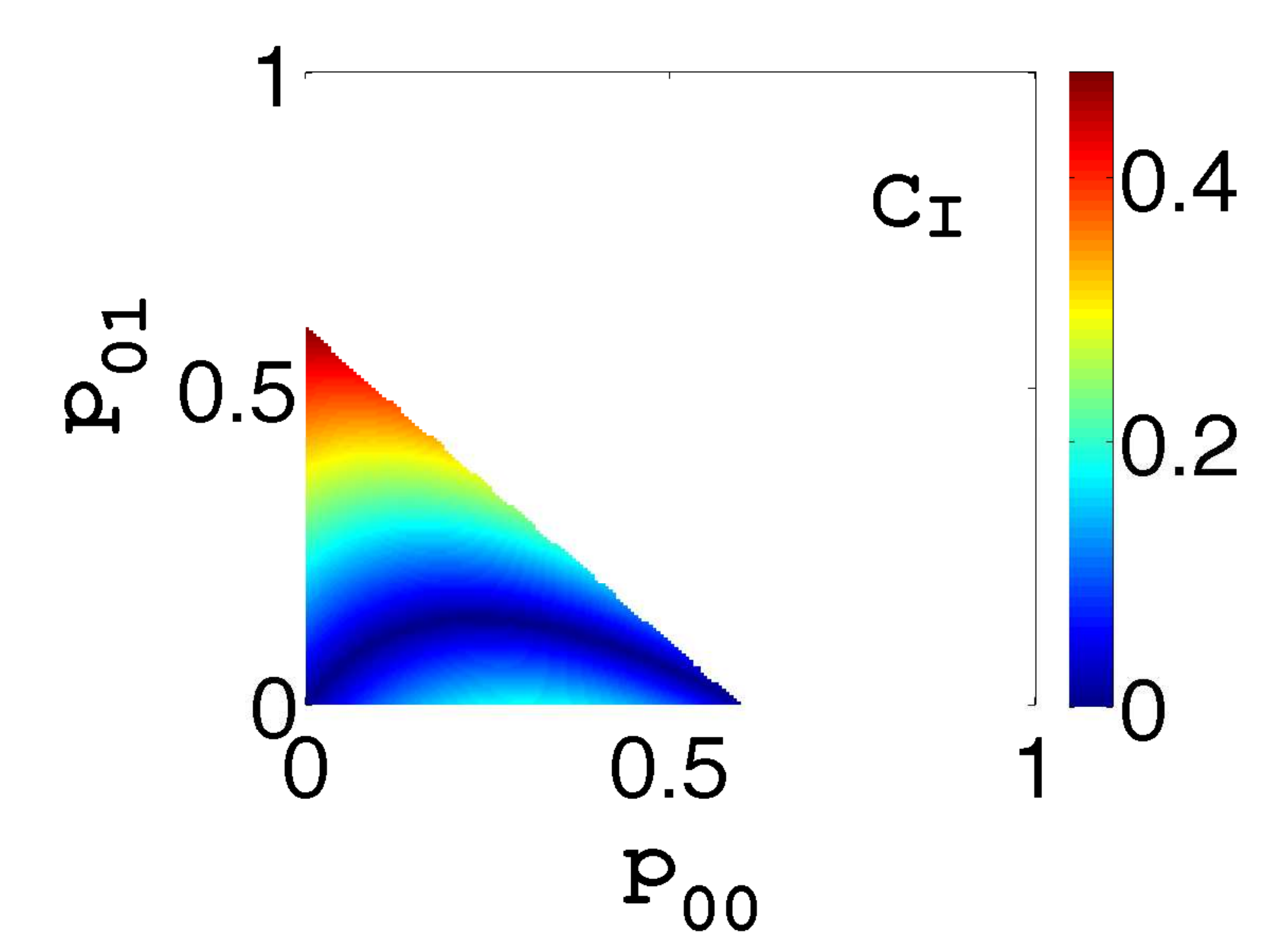}
\label{fig2:2:1} } \subfloat[$C_{II}$ when $p_{11}=0.4$.]{
\includegraphics[width=0.24\textwidth]{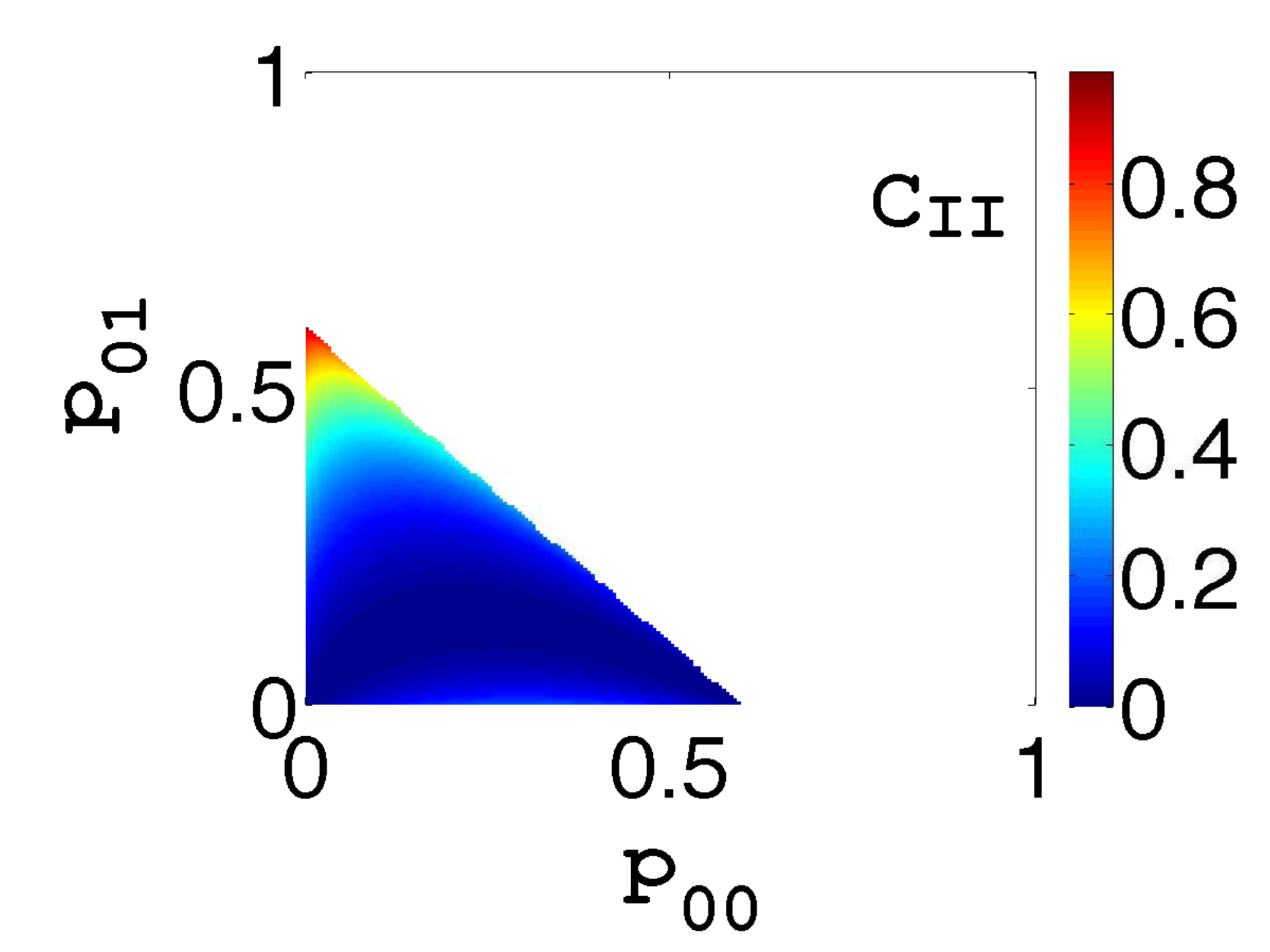}
\label{fig2:2:2} } \\
 \subfloat[$C_{I}$ when $p_{11}=0.7$.]{ \includegraphics[width=0.24\textwidth]{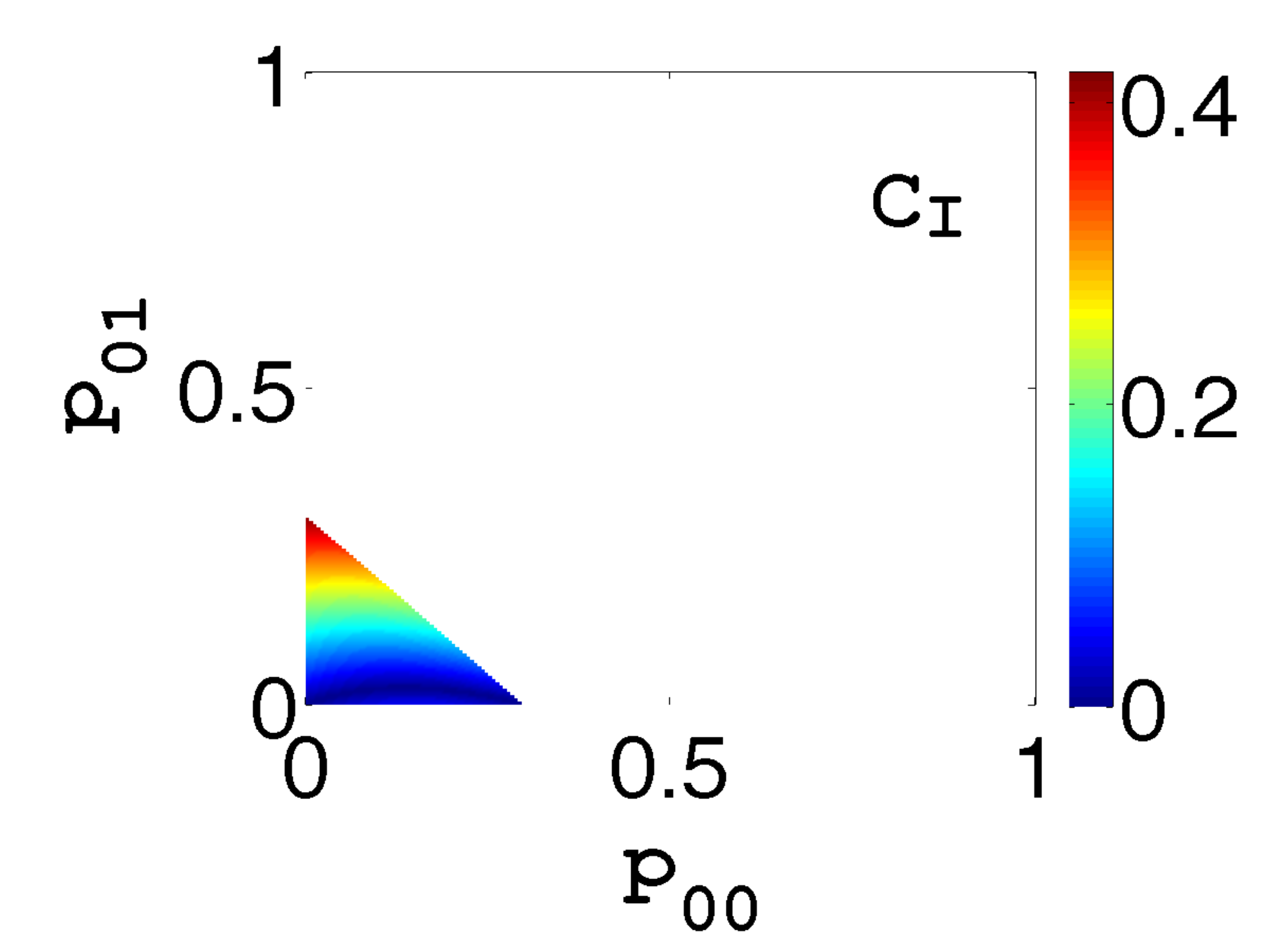}
\label{fig2:3:1} } \subfloat[$C_{II}$ when $p_{11}=0.7$.]{
\includegraphics[width=0.24\textwidth]{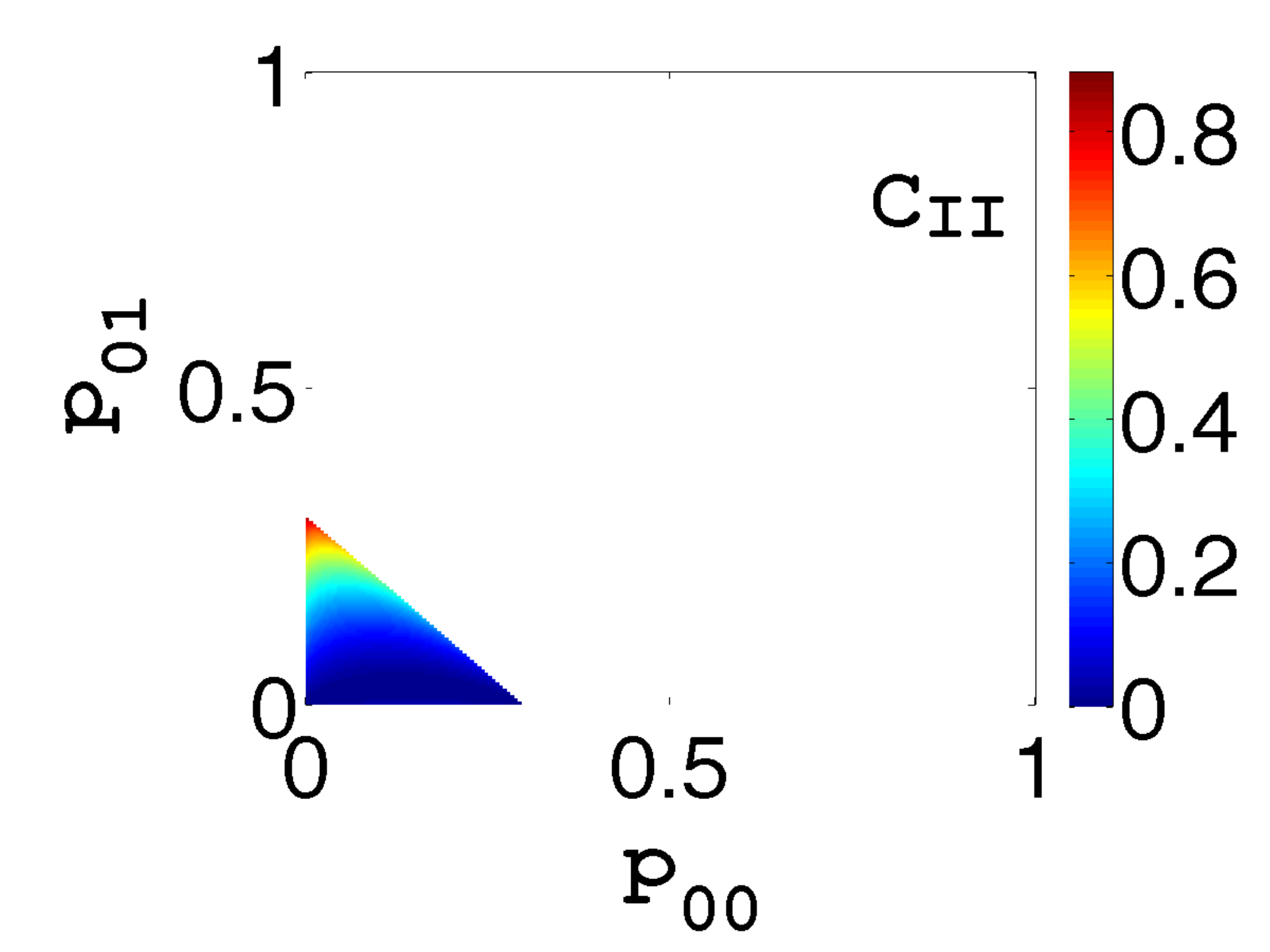}
\label{fig2:3:2} } \caption{Numerical comparison of $C_{I}$ and $C_{II}$ with varying $p_{11}$.}

\label{fig2} 
\end{figure}

The numerical results in Fig. \ref{fig1} and Fig. \ref{fig2} show
the similarities of these two measures, which supports that they both
characterize the same quantity --- the degree of correlation.

\subsection{Analytical results}

The above numerical results seem to support the validity of Eq. (\ref{c1c2so}).
However, we will prove the following proposition: \textit{ there exist
two quantum states such that Eq. (\ref{c1c2so}) is no longer true.}

Let us construct two quantum states that do not satisfy Eq. (\ref{c1c2so}).

\begin{figure}[ht]
 \centering \includegraphics[width=0.45\textwidth]{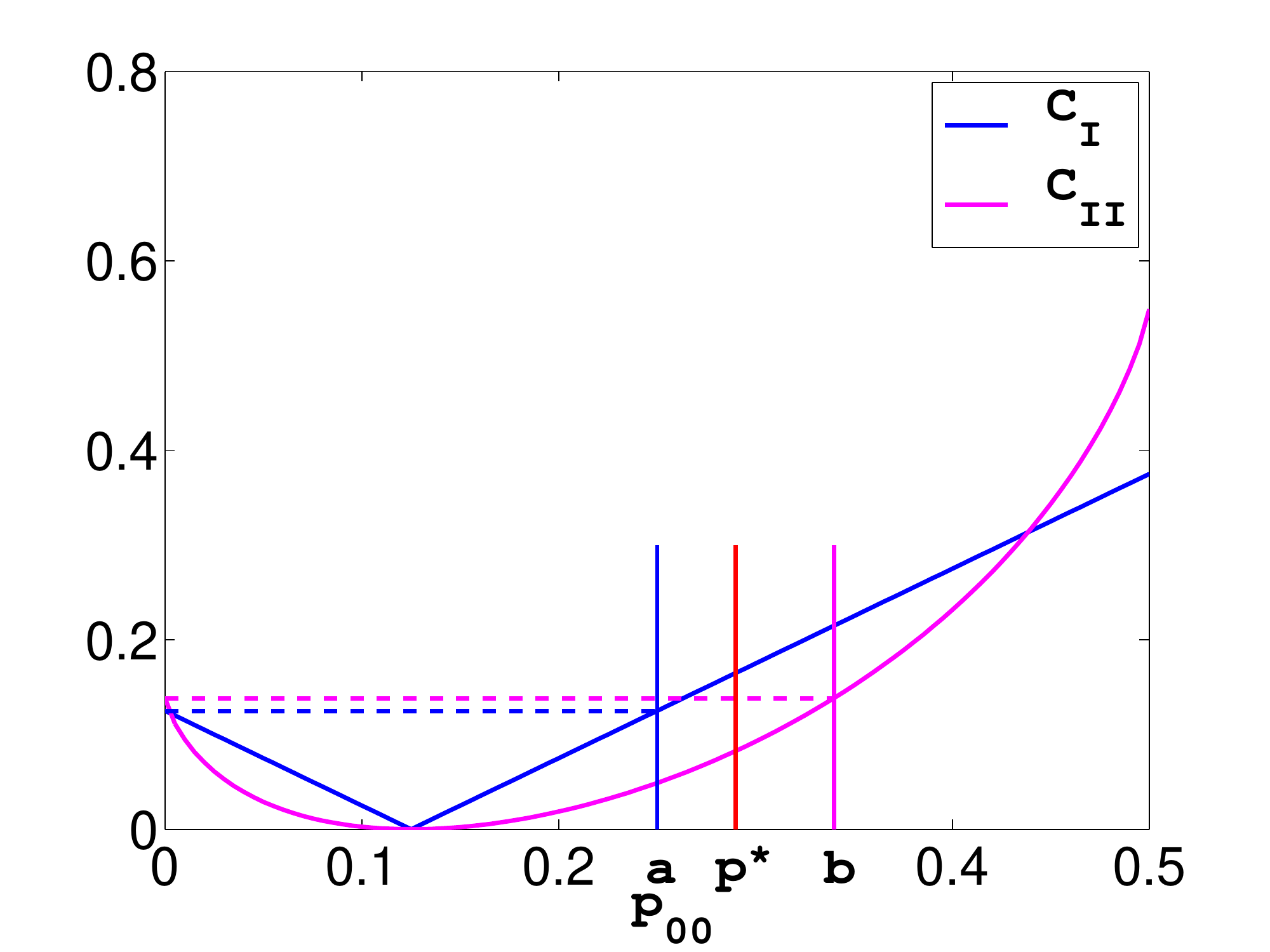}
\caption{$C_{I}$ and $C_{II}$ varying with $p_{00}$ when $p_{10}=\frac{1}{8}$
and $p_{11}=\frac{3}{8}$.}

\label{fig3} 
\end{figure}

We take $p_{10}=\frac{1}{8}$ and $p_{11}=\frac{3}{8}$ in the quantum
state (\ref{rhop}). Then $p_{00}\in[0,\frac{1}{2}]$, and Eq. (\ref{c1p})
and (\ref{c2p}) become \begin{equation}
C_{I}(p_{00})=\vert p_{00}-\frac{1}{8}\vert,\end{equation}
 and \begin{eqnarray}
C_{II}(p_{00}) & = & p_{00}\ln p_{00}+(\frac{1}{2}-p_{00})\ln(\frac{1}{2}-p_{00})\nonumber \\
 &  & -(\frac{1}{8}+p_{00})\ln(\frac{1}{8}+p_{00})-(\frac{7}{8}-p_{00})\nonumber \\
 &  & \times\ln(\frac{7}{8}-p_{00})+\frac{3\ln3-4}{8},\end{eqnarray}
 which are demonstrated in Fig. \ref{fig3}. %

A simple calculation gives that $C_{I}(0)=\frac{1}{8}$, $C_{I}(\frac{1}{8})=0$,
$C_{I}(\frac{1}{2})=\frac{3}{8}$, $C_{II}(0)=2+\frac{3\ln3-7\ln7}{8}$,
$C_{II}(\frac{1}{8})=0$, and $C_{II}(\frac{1}{2})=2-\frac{5\ln5}{8}$.
It is easy to check that $C_{I}(\frac{1}{8})<C_{I}(0)<C_{I}(\frac{1}{2})$
and $C_{II}(\frac{1}{8})<C_{II}(0)<C_{II}(\frac{1}{2})$. We can also
prove that $C_{I}(p_{00})$ and $C_{II}(p_{00})$ are continuous strictly
increasing function when $p_{00}\in[\frac{1}{8},\frac{1}{2}]$. According
to the intermediate value theorem, there exist $a,b\in[\frac{1}{8},\frac{1}{2}]$,
such that $C_{I}(a)=C_{I}(0)$ and $C_{II}(b)=C_{II}(0)$.

We will prove that $a<b$. It is easy to observe that $C_{I}(\frac{1}{4})=C_{I}(0)=\frac{1}{8}$,
so $a=\frac{1}{4}$. A direct calculation gives \[
C_{II}(\frac{1}{4})-C_{II}(0)=\frac{1}{8}\ln\frac{823543}{1350000}<0,\]
 which implies that $C_{II}(\frac{1}{4})<C_{II}(0)=C_{II}(b)$. Since
the function $C_{II}$ is strictly increasing function when $p_{00}\in[\frac{1}{8},\frac{1}{2}]$,
we have $a=\frac{1}{4}<b$.

$\forall p^{\star}\in(a,b)$, we have \begin{eqnarray}
 &  & C_{I}(0)=C_{I}(a)<C_{I}(p^{\star}),\label{ineqa}\\
 &  & C_{II}(0)=C_{II}(b)>C_{II}(p^{\star}).\label{ineqb}\end{eqnarray}
 In Eqs. (\ref{ineqa}) and (\ref{ineqb}), we use the fact again
that $C_{I}$ and $C_{II}$ are strictly increasing functions when
$p_{00}\in[\frac{1}{8},\frac{1}{2}]$. This completes our proof of
the proposition.

\section{Discussions}

\begin{figure}[ht]
\centering \includegraphics[width=0.45\textwidth]{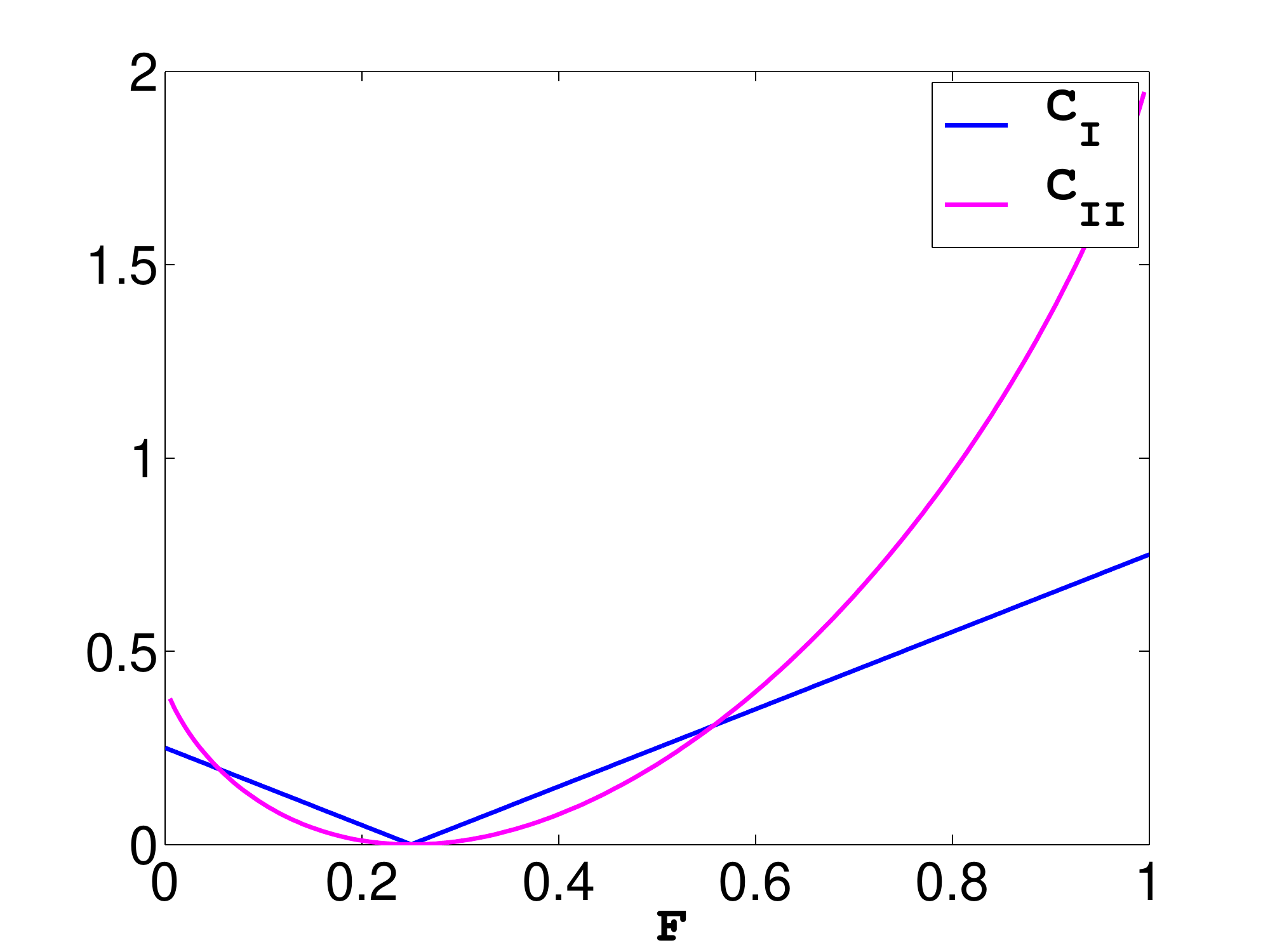}
\caption{The two correlation measures for the Werner states.}

\label{fig4} 
\end{figure}

First, we will address one aspect in the relations between correlation
and entanglement. Since entanglement is a kind of quantum correlation,
it is interesting to ask whether the appearance of entanglement can
be identified in the degree of correlation. Here we will try to give
an answer with the Werner states \cite{Wer89}, which are defined
by \begin{equation}
\rho_{W}(F)=\frac{1-F}{3}I^{(1)}I^{(2)}+\frac{4F-1}{3}\vert\Psi^{-}\rangle\langle\Psi^{-}\vert,\end{equation}
 where $0\le F\le1$ and $\vert\Psi^{-}\rangle=\frac{1}{\sqrt{2}}(\vert01\rangle-\vert10\rangle)$.
As is well known, the Werner states are entangled if and only if $F>\frac{1}{2}$
\cite{BDSW96}. The expressions for the two correlation measures are
given respectively by \begin{equation}
C_{I}(\rho_{W}(F))=\vert F-\frac{1}{4}\vert,\end{equation}
 and \begin{equation}
C_{II}(\rho_{W}(F))=\ln\frac{4}{3}+F\ln 3+F\ln F+(1-F)\ln(1-F),\end{equation}
 which are numerically demonstrated in Fig. \ref{fig4}. %

As shown in Fig. \ref{fig4}, there is not any evidence in the two
correlation measures that shows the appearance of entanglement at
$F=\frac{1}{2}$.

Second, the two correlation measures are related by the following
inequalities \cite{NC00,OP93,WVHC08}: \begin{equation}
2C_{I}^{2}(\rho^{(12)})\le C_{II}(\rho^{(12)})\le2C_{I}(\rho^{(12)})\ln d+\frac{1}{e},\label{ineq}\end{equation}
where $d$ is the dimension of the Hilbert space of the composite system.
 When $2C_{I}(\rho^{(12)})\le\frac{1}{e}$, the right part of the
inequality (\ref{ineq}) can be strengthened to \begin{equation}
C_{II}(\rho^{(12)})\le2C_{I}(\rho^{(12)})\ln d-2C_{I}(\rho^{(12)})\ln(2C_{I}(\rho^{(12)})).\label{ineq2}\end{equation}
 The inequalities (\ref{ineq}) and (\ref{ineq2}) show that we can
estimate the range of values taken by $C_{II}$ from the value of
$C_{I}$.

Third, in addition to the two correlation measures $C_{I}$ and $C_{II}$,
there exist other correlation measures, e.g. \begin{equation}
C_{III}(\rho^{(12)})=A(\rho^{(12)},\tilde{\rho}^{(12)}),\end{equation}
 where the angle distance $A(\sigma,\tau)\equiv\arccos F(\sigma,\tau)$
with the fidelity $F(\sigma,\tau)\equiv\mathrm{Tr}\sqrt{\sigma^{\frac{1}{2}}\tau\sigma^{\frac{1}{2}}}$
\cite{NC00}. In this measure, the fidelity $F(\rho^{(12)},\tilde{\rho}^{(12)})$
plays a central role. In fact, we can construct a correlation measure
directly from the Fidelity by \begin{equation}
C_{III}^{\prime}(\rho^{(12)})=1-F^{2}(\rho^{(12)},\tilde{\rho}^{(12)}).\end{equation}
 It is easy to prove that $C_{III}(\rho^{(12)})$ and $C_{III}^{\prime}(\rho^{(12)})$
will give the same order on the degree of correlation for any two
partite quantum states.

\section{Summary}

We briefly review two different correlation measures for a bipartite
quantum state, and associate the first correlation measure with the
correlation functions. Comparing these two correlation measures for
the classically correlated two-qubit states, we observe that they
give the same ordering on the degrees of correlation for most quantum
states. However, we find that the two correlation measures can give different
orderings on the degrees of correlations for some two specific two-qubit states. This situation is
similar to the work on entanglement monotone, where different entanglement
monotones may give different orderings of two quantum states on the
degree of entanglement. Our work present a comparison between the
correlation functions and mutual information on characterizing the
degrees of correlation in multipartite quantum states , which sheds a novel light on the interplay
between many-body physics and quantum information science.

\hspace{1mm}

\noindent\textbf{Acknowledgments}:
This work is supported by NSF of China under Grant No. 10775176 and
10975181, and NKBRSF of China under Grants No. 2006CB921206.

\bibliographystyle{apsrev4-1}
\bibliography{cfmiref.bib}

\end{document}